\documentclass{PoS}

\usepackage{amsmath}
\usepackage{subfigure}
\usepackage{braket}

\newcommand{\be}{\begin{equation}}
\newcommand{\ee}{\end{equation}}
\newcommand{\bea}{\begin{eqnarray}}
\newcommand{\eea}{\end{eqnarray}}

\title{Charm hadrons and lattice QCD}

\ShortTitle{Charm hadrons and lattice QCD}

\author{\speaker{Lorenzo Riggio}\\
        Istituto Nazionale di Fisica Nucleare, Sezione di Roma Tre, Rome, Italy\\
        E-mail: \email{lorenzo.riggio@gmail.com}}

\author{Giorgio Salerno\\
        Universit\'a Roma Tre \& Istituto Nazionale di Fisica Nucleare, Sezione di Roma Tre, Rome, Italy\\
        E-mail: \email{salerno@fis.uniroma3.it}
        \\
        \textbf{for the ETM Collaboration}}

\abstract{We review two lattice calculations involving charm hadrons: a determination of the vector and scalar form factors of the semileptonic $D \to \pi \ell \nu$ decays, which are relevant for the extraction of the CKM matrix element $\lvert V_{cd} \rvert$ from experimental data, and a calculation of the matrix elements of four-fermion operators relevant to the description of the neutral D mixing in the Standard Model and its extensions. Both analyses are based on the gauge configurations produced by the European Twisted Mass Collaboration with $N_f = 2 + 1 + 1$ flavors of dynamical quarks. We simulated at three different values of the lattice spacing and with pion masses as small as 210 MeV. }

\FullConference{VIII International Workshop On Charm Physics\\
                 5-9 September, 2016\\
                 Bologna, Italy}

\begin{document}

\section{Introduction and simulation details}

Flavor Physics offers a unique possibility for an indirect discovery of New Physics (NP) effects through virtual exchanges of yet-to-be-discovered heavy particles in loop suppressed processes. This approach, which is particularly promising for processes that are highly suppressed within the Standard Model (SM), proved to be very successful in the past allowing for the indirect determination of the charm and top quark mass~\cite{Glashow:1970gm, Gaillard:1974hs, Albrecht:1987dr}.

Moreover Flavor Physics data play a major role in providing stringent tests of the Cabibbo-Kobayashi-Maskawa (CKM) matrix \cite{CKM} which, in the SM, parametrises the relative strength of different flavour-changing weak processes. Inconsistencies in the CKM-picture would indicate the presence of new physics beyond the SM. Even if all the precision tests of the SM performed so far remain in agreement with the CKM paradigm, the absence of deviations provides stringent constraints on nonstandard phenomena and their energy scale. It is therefore important to determine all CKM-matrix elements as precisely as possible by studying flavour changing processes both experimentally and theoretically.

In this scenario Lattice QCD has a primary role as it allows to compute, non-perturbatively and from first principles, hadronic quantities such as decay constants, form factors and bag parameters. Thanks to the astonishing progress in algorithms and machines Lattice QCD has entered the precision era as the accuracy of numerical computations is becoming comparable to that of experiments. For some of the relevant hadronic quantities in Flavour Physics the goal of per cent precision has been achieved. State-of-the-art lattice calculations involve $\mathcal{O}(a)$-improved fermionic actions with $N_f =2+1+1$ dynamical flavors (which include in the sea, besides two light mass-degenerate quarks, also the strange and the charm quarks \cite{Baron:2010bv,Baron:2011sf}), with the smallest simulated pion masses being today at the physical point or slightly higher, and employing three or more values of the lattice spacing. Many recent calculations simulate the charm quark around its physical value. 
This has become possible thanks to the increasing availability of dynamical gauge field ensembles with fine lattice spacings and nowadays direct computations by many lattice collaborations have shown that the cutoff effects in the D-sector are small and under control.

In this contribution we present two lattice QCD calculations involving charm hadrons based on the gauge ensembles produced by the European Twisted Mass Collaboration (ETMC).  After motivating lattice efforts in the charm sector, we will present in section \ref{sec:Dtopi} our preliminary determination of the vector and scalar form factors of the semileptonic $D \to \pi \ell \nu$ decays, as functions of the squared 4-momentum transfer $q^2$. Afterwards, in section \ref{sec:DDbar}, we will present a determination of the bag-parameters relevant for the description of the $\Delta C=2$ transitions occurring in $\bar{D}^0-D^0$ oscillations.
A detailed description of the methodology followed in this two analyses is beyond the scope of this contribution, we will rather summarize the main features and results and point out the differences with respect to other lattice calculations. We refer the interested reader respectively to Ref. \cite{Lubicz:2016wwx} and \cite{Carrasco:2015pra} for the technical details of the numerical work.

The simulations were carried out at various lattice volumes and for three values of the lattice spacing in the range $0.06 - 0.09$ fm in order to keep under control the  extrapolation to the continuum limit. 
The simulated pion masses range from $\approx 210$ MeV to $\approx 450$ MeV. 
The gauge fields were simulated using the Iwasaki gluon action \cite{Iwasaki:1985we}, while sea and light valence quarks were implemented with the Wilson Twisted Mass Action \cite{Frezzotti:2003xj,Frezzotti:2003ni}.
The valence charm quark was simulated using the Osterwalder-Seiler action \cite{Osterwalder:1977pc}.
At maximal twist such a setup guarantees the automatic $\mathcal{O}(a)$-improvement \cite{Frezzotti:2003ni,Frezzotti:2004wz}.
More details about the lattice ensembles and the simulation details can be found in Ref.~\cite{Carrasco:2014cwa}.

\section{Lorentz symmetry breaking in the behavior of the scalar and vector form factors}
\label{sec:Dtopi}

A lattice calculation of the vector and scalar form factors of the semileptonic $D \to \pi \ell \nu$ decays, combined with experimental data, enable determinations of the CKM matrix element $\lvert V_{cd} \rvert$. In fact the relation between semileptonic $D \to \pi \ell \nu$ decay width, with $\ell=e, \mu$, and the matrix element $\lvert V_{cd} \rvert$, up to well known overall factors, is
\begin{equation}
 \frac{{d\Gamma \left( {D \to \pi l\nu } \right)}}{{d{q^2}}} \propto {\lvert {{V_{cd}}} \rvert^2}{\lvert {{f_ +^{D\pi }} \left( {{q^2}} \right)} \rvert^2}.
 \end{equation} 

In practice, most lattice-QCD calculations of $D \to \pi \ell \nu$ focus on providing the value of the vector form factor at a single value of the squared 
4-momentum transfer, ${f_ +^{D\pi }} \left( {{q^2=0}} \right)$, which is sufficient to obtain $\lvert V_{cd} \rvert$. Our calculation also provides a determination 
of the $D \to \pi \ell \nu$ form factors in the whole kinematical range of values of the squared 4-momentum transfer $q^2$ accessible in the experiments,
 i.e.~from $q^2 = 0$ to $q^2 = \left( M_D - M_\pi \right)^2$, thereby allowing a comparison of the shapes of the lattice simulation and experimental data.

Momenta were injected on the lattice using non-periodic boundary conditions for the quark fields \cite{Bedaque:2004kc,deDivitiis:2004kq,Guadagnoli:2005be}, obtaining in this way values ranging from $\approx 150$ MeV up to $\approx 650$ MeV.  
The matrix elements of both vector and scalar currents are determined for a plenty of  kinematical conditions in which parent and child mesons are either moving or at rest. 
Employing in the simulation different values of the lattice spacing $a$ allows for a controlled continuum limit in which unphysical lattice artifact are removed from the quantity of interest. Moreover, $\mathcal{O}(a)$-improvement in the lattice action guaranties that the quantity calculated on the lattice and the physical one can only be different up to $a^2$-terms. This is a standard procedure of every lattice calculation. However in our analisys of the $D \to \pi \ell \nu$ semileptonic form factors the impact of the discretization effects manifested itself in a more subtle way. Lorentz symmetry breaking due to hypercubic effects has been clearly observed in the data (see also Ref.~\cite{Carrasco:2015bhi}). 
In this contribution we also present the removal of such hypercubic effects and the determination of the physical, Lorentz-invariant, semileptonic vector and scalar form factors.

The vector semileptonic form factor $f_+(q^2)$ and the scalar semileptonic form factor $f_0(q^2)$ are related to matrix elements corresponding to the time and spacial components of the weak vector current $V_\mu = \bar{c} \gamma_\mu d$ via
 \bea
     \label{Vmu_L}
      \left\langle {\pi \left( {{p_\pi }} \right)} \right|\;{{\hat V}_\mu }\;\left| {D\left( {{p_D}} \right)} \right\rangle  = {P_\mu }{f_ + }\left( {{q^2}} \right) + {q_\mu }{f_ - }\left( {{q^2}} \right)~ , \\
     \label{f0def}
     {f_0}\left( {{q^2}} \right) = {f_ + }\left( {{q^2}} \right) + \frac{{{q^2}}}{{{M}_D^2 - {M}_\pi ^2 }}{f_ - }\left( {{q^2}} \right) ~ ,
 \eea
 where $P \equiv p_D + p_\pi$, $q \equiv p_D - p_\pi$. The scalar form factor is also related, via Ward-Takahashi identity, to the matrix element of the scalar current $V_S =  \bar{c} d$ through
\be
     \label{S_L}
     \braket{\pi(p_\pi)| \hat{V}_S |D(p_D)} =  f_0(q^2) (M_D^2 - M_\pi^2) / (m_c - m_\ell) ~ ,
\ee 
where $m_{\ell(c)}$ are the renormalized light (charm) quark masses. In Eqs.~(\ref{Vmu_L}-\ref{S_L}) $\hat{V}_\mu$ and $\hat{V}_S$ indicate the renormalized vector and scalar currents.
Once these matrix elements have been computed on the lattice for each gauge ensemble and each choice of parent and child meson momenta, $\vec{p}_D$ and $\vec{p}_\pi$, the vector and scalar form factors can be determined as best-fit values of the set of the matrix elements corresponding to the time and spacial components of the vector current and to the scalar current.
 
The matrix elements of the weak vector and scalar currents can be extracted from the large time distance behavior of a convenient combination of Euclidean 3-points and 2-points correlation functions in lattice QCD. 
As it is well known, at large time distances 2- and 3-points correlation functions behave as
 \be
      \widetilde{C}_2^M(t) \equiv \frac{1}{2}\left[ C_2^M(t) + \sqrt{C_2^M(t)^2 - C_2^M(T/2)^2} \right] ~ _{\overrightarrow{t \gg a}} ~ Z_M(\vec{p}_M)\, e^{-E_M t} / (2 E_M) ~ , 
     \label{C2}
 \ee
 \be
     C^{D \pi}_{\mu,S}(t, t_s) ~ _{\overrightarrow{t \gg a, (t_s - t) \gg a}} ~ \sqrt{Z_D(\vec{p}_D) Z_\pi(\vec{p}_\pi)} ~ \braket{\pi(p_\pi) | V_{\mu,S} | D(p_D)} ~ 
                                            e^{-E_D t - E_\pi (t_s - t)} / (4 E_D E_\pi) ,
     \label{C3}
\ee
where $M$ stands for either the $D$ or the $\pi$ meson, $E_M$ is the meson energy, $t_s$ is the time distance between the source and the sink, ${Z_{\,D}}(\vec{p}_D)$ and ${Z_\pi}(\vec{p}_{\pi})$ are the matrix elements $\lvert \braket{D(\vec{p}_D) \lvert\,\bar{c}\,\gamma_5\,d\,\rvert\,0}\rvert^2$ and $\lvert \braket{\pi(\vec{p}_{\pi})\lvert\,\bar{u}\,\gamma_5\,d\,\rvert\,0}\rvert^2$, where the dependence on the meson momenta $\vec{p}_D$ and $\vec{p}_\pi$ arises from the use of smeared interpolating fields.

The correlation functions can be combined in the five ratios $R_\mu$ ($\mu = 0, 1, 2, 3$) and $R_S$ as
 \be
    R_\mu(t) \equiv 4 p_{D \mu} p_{\pi \mu} \frac{C^{D\pi}_\mu(t, t_s, \vec{p}_D, \vec{p}_\pi) C^{\pi D}_\mu(t, t_s, \vec{p}_\pi, \vec{p}_D)}
        {C^{\pi \pi}_\mu(t, t_s, \vec{p}_\pi, \vec{p}_\pi) C^{DD}_\mu(t, t_s, \vec{p}_D, \vec{p}_D)} ~ _{\overrightarrow{t \gg a}} ~ 
        \left( \braket{\pi(p_\pi) | \hat{V}_\mu |D(p_D)} \right)^2 ,
    \label{Rmu}
 \ee
 \be
    R_S(t) \equiv 4 E_D E_\pi \left( \frac{\mu_c - \mu_\ell}{M_D^2 - M_\pi^2} \right)^2  \frac{C^{D\pi}_S(t, t_s) C^{\pi D}_S(t, t_s)} {\widetilde{C}_2^D(t_s) \widetilde{C}_2^\pi(t_s)} 
        ~ _{\overrightarrow{t \gg a}} ~ \left( \frac{m_c - m_\ell}{M_D^2 - M_\pi^2} \braket{\pi(p_\pi)| \hat{V}_S |D(p_D)} \right)^2 ~ , 
    \label{RS}
 \ee
where $\mu_{\ell(c)}$ are the bare light (charm) quark masses and $m_{\ell(c)}$ are the corresponding renormalized ones. 
In the r.h.s.~of Eqs.~(\ref{Rmu}-\ref{RS}) $\hat{V}_\mu$ and $\hat{V}_S$ are already the renormalized vector and scalar currents, respectively, since the multiplicative RCs $Z_V$ and $Z_P$ cancel out in the ratios.
The matrix elements of the vector and scalar currents can therefore be easily extracted from the plateaux of $R_\mu(t)$ and $R_S(t)$ at large time distances.

\begin{figure}[htb!]
\centering
\includegraphics[scale=0.30]{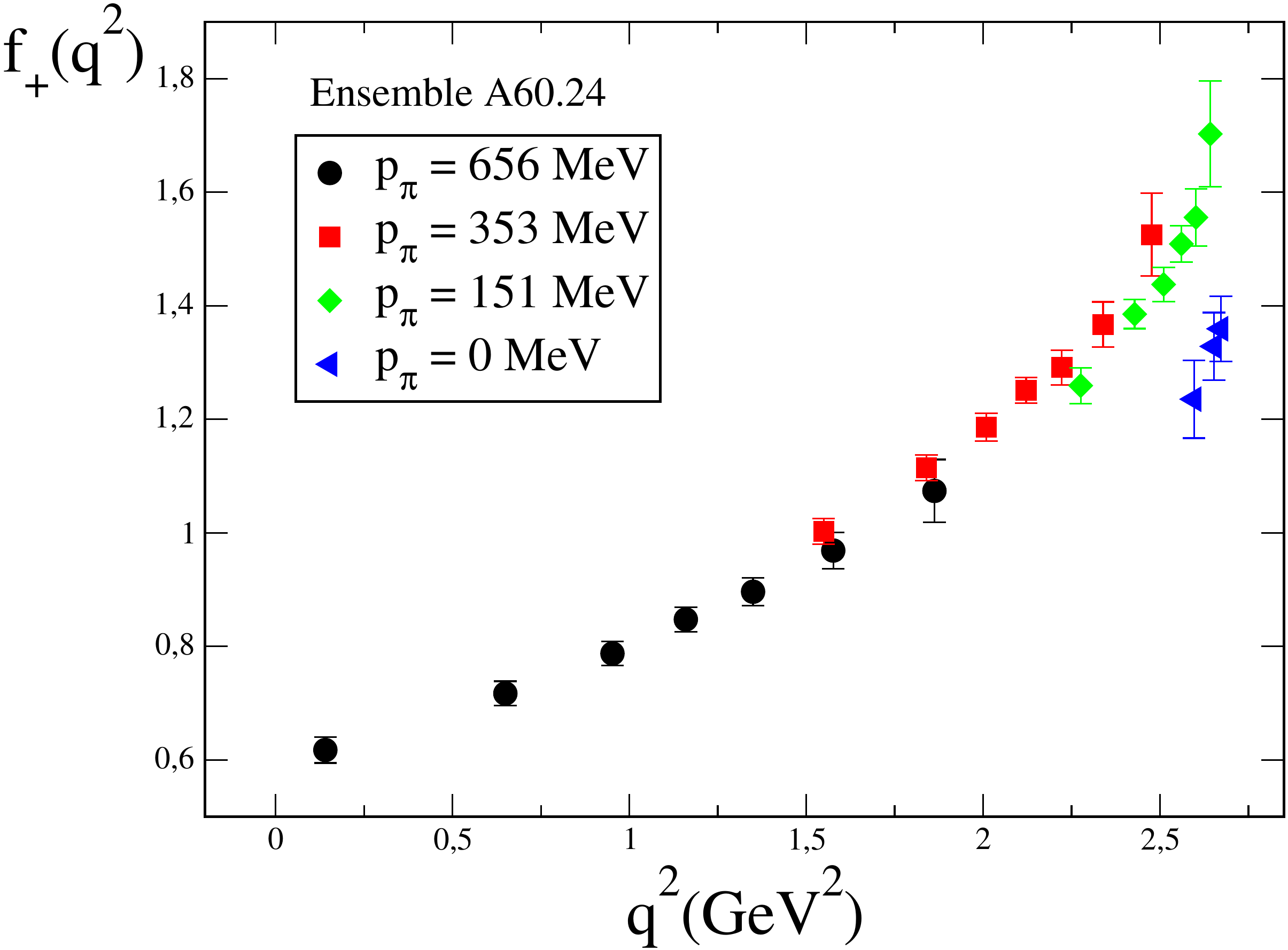}
\includegraphics[scale=0.30]{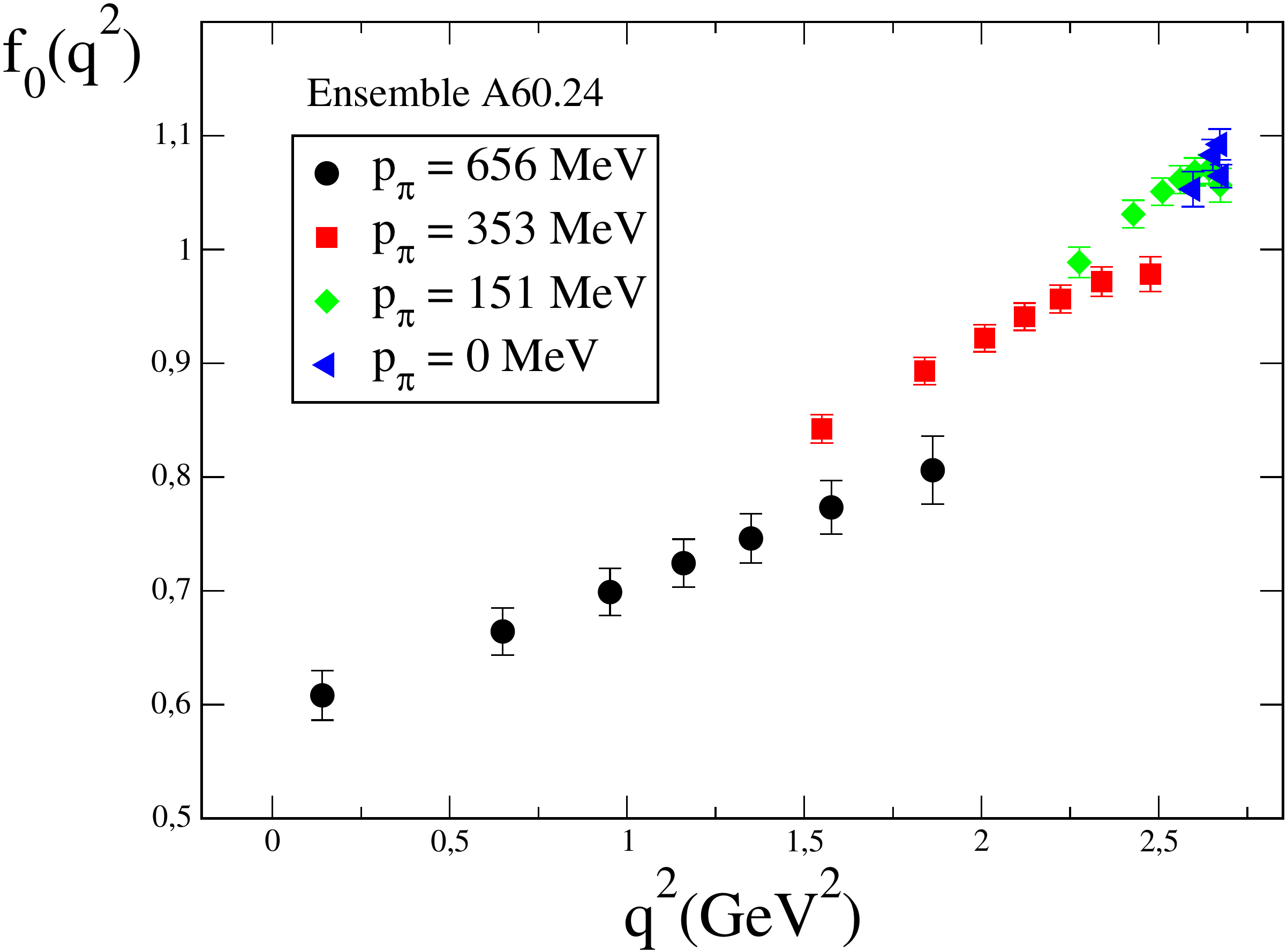}
\caption{\it \footnotesize Momentum dependence of the vector (left panel) and scalar (right panel) form factors in the case of the gauge ensemble A60.24 \protect\cite{Carrasco:2014cwa}. Different markers and colors distinguish different values of the child meson (pion) momentum. The simulated pion mass is $M_\pi \simeq 390$ MeV and the charm quark mass corresponds to its physical value $m_c^{phys}$ from \protect\cite{Carrasco:2014cwa}.}
\label{fishbone}
\end{figure}

The results, interpolated at the physical value of the charm quark mass $m_c^{phys}$ determined in \cite{Carrasco:2014cwa}, are shown in Fig.~\ref{fishbone} in the case of the gauge ensemble A60.24 (see Ref.~\cite{Carrasco:2014cwa}). 
It can be clearly seen that the extracted form factors are not Lorentz-invariant quantity since a dependence on the value of the child meson momentum is visible.

The physical form factors can only depend on the squared 4-momentum transfer $q^2$, which is a Lorentz-invariant quantity, while other dependecies are forbidden by the space-time symmetries. On the lattice however, $q^2$ is no longer the sole invariant. Since the lattice is only invariant under discrete $90^\circ$  rotations, beside $q^2$, other hypercubic invariants may appear. These effects must be proportional to the lattice spacing $a$ so that, in the continuum limit, the correct  $q^2$-dependence is recovered. Therefore, since in our setup all the current matrix elements are $O(a)$-improved \cite{Carrasco:2016kpy} we tried to describe the breaking of the Lorentz invariance by means of $O(a^2)$ hypercubic effects. In principle, the behavior observed in Fig.~\ref{fishbone} might be also (at least partially) related to finite volume effects \cite{Bijnens:2014yya} but a direct comparison of two gauge ensembles, A40.24 and A40.32, which share the same pion mass and lattice spacing at different lattice sizes, $L = 24 a$ and $L = 32 a$ (see Ref.~\cite{Carrasco:2014cwa}), suggests that finite size hypercubic effects are negligible.

In the case of vector current matrix elements possible hypercubic terms have to be odd in the meson momenta and therefore at order $O(a^2)$ we can add to the Lorentz-covariant decomposition (\ref{Vmu_L}) the following hypercubic structure 
 \be
    \braket{\pi(p_\pi) | \hat{V}_\mu^{hyp} |D(p_D)}  = a^2 \left[ q_\mu^3 H_1+ q_\mu^2 P_\mu H_2 + q_\mu P_\mu^2 H_3 + P_\mu^3 H_4 \right] ~ ,
    \label{VH}
 \ee
where $H_i$ ($i=1,...,4$) are additional form factors.
We adopt for them a simple polynomial form in the $z$ variable \cite{Bourrely:2008za}, which depends on both $q^2$ and $m_\ell$, viz.
 \be
    H_i(z) = d_0^i + d_1^i z + d_2^i z^2 ~ ,
    \label{hyp_form_factors}
 \ee
where $d_{0,1,2}^i$ will be treated as free parameters in the fitting procedure.

In a similar way one can consider the possible presence of $O(a^2)$ hypercubic terms in the scalar matrix elements.
The Ward-Takahashi identity (WTI), relating the 4-divergence of the vector current to the scalar density, is a good place to look for such hypercubic terms.
We have indeed observed WTI violations that cannot be interpreted as $a^2$ and/or $a^2 q^2$ (Lorentz-invariant) discretization effects. A direct investigation of our data (see~\cite{Lubicz:2016wwx}) suggests a simple linear dependence on the hypercubic invariant $q^{[4]} \equiv \sum_\mu q_\mu^4$.
Thus we have considered the presence of $O(a^2)$ hypercubic effects in the WTI in the form
 \be
    q^\mu \braket{\pi(p_\pi) | \hat{V}_\mu | D(p_D)} = \left(m_c - m_\ell \right) \braket{\pi(p_\pi) | \hat{V}_S | D(p_D)} +a^2 ~ q^{[4]} ~ H_5 ~ ,
    \label{WI}
 \ee
where for $H_5$ we have assumed the simple form $H_5 = d_0^5 + d_1^5 m_\ell$ with $d_{0,1}^5$ being free parameters.

As for the Lorentz-invariant form factors $f_{+,0}(q^2)$ appearing in Eqs.~(\ref{Vmu_L}-\ref{S_L}) we adopt the modified z-expansion of Ref.~\cite{Bourrely:2008za} and impose the condition $f_+(0) = f_0(0) = f(0)$ (as in Ref.~\cite{Carrasco:2016kpy}):
 \be
    f_{+(0)}(q^2) = \frac{f(0) + c_{+(0)} (z - z_0) \left(1 + \frac{z + z_0}{2} \right)}{1 - K_{FSE}^{+(0)} ~ q^2  / M_{V(S)}^2} ~ , 
   \label{z-exp}
 \ee
where $z_0 \equiv z(q^2 = 0$).
Inspired by the hard pion SU(2) ChPT \cite{Bijnens:2010jg}, $f(0)$ can be written as
 \be
    f(0) = b_0 \left[ 1 - \frac{3}{4} \left( 1 + 3 g^2 \right) ~ \xi_\ell \log\xi_\ell + b_1 ~ \xi_\ell + b_2 ~ a^2 \right]~ ,  
    \label{ChLim}
 \ee
with $\xi_\ell = 2B m_\ell/ (16\pi^2f^2)$, where $B$ and $f$, determined in \cite{Carrasco:2014cwa}, are the SU(2) low-energy constants entering the LO chiral Lagrangian. 
The coefficients $b_i$ ($i = 0, 1, 2$) are treated as free parameters in the fitting procedure, while $g$ is kept constant at the value $g = 0.61$ \cite{PDG}.
In Eq.~(\ref{z-exp}) the quantity $M_{V(S)}$, representing the vector (scalar) pole mass, is treated as a free parameter, while for the coefficients $c_{+(0)}$ we assume a simple linear dependence on $a^2$.
Finally the term $K_{FSE}^{+(0)}$ takes into account finite size effects (FSE) which on the lattice are typically proportional to $e^{-M_\pi L}$.

\begin{figure}[htb!]
\centering
\includegraphics[scale=0.30]{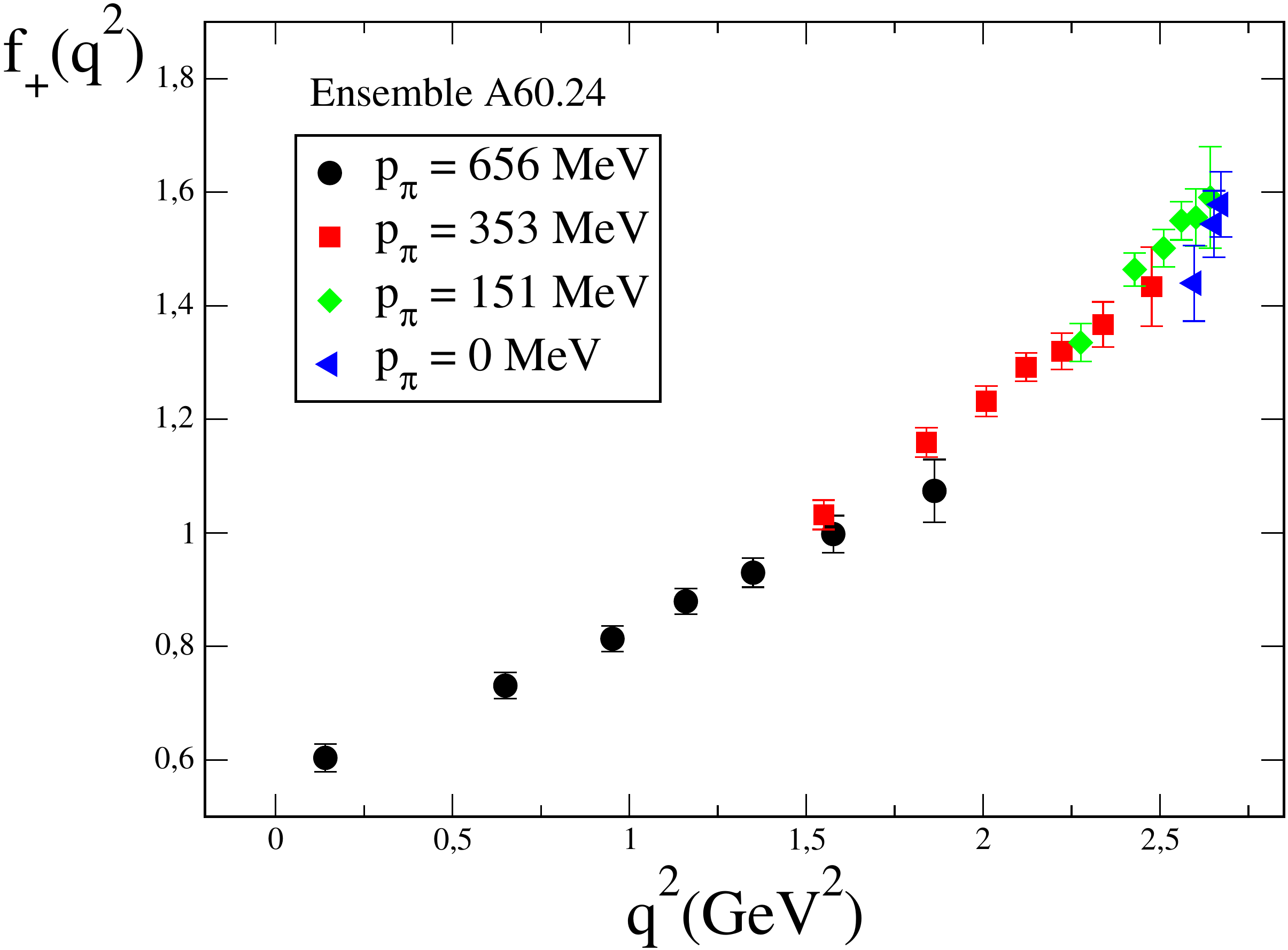}
\includegraphics[scale=0.30]{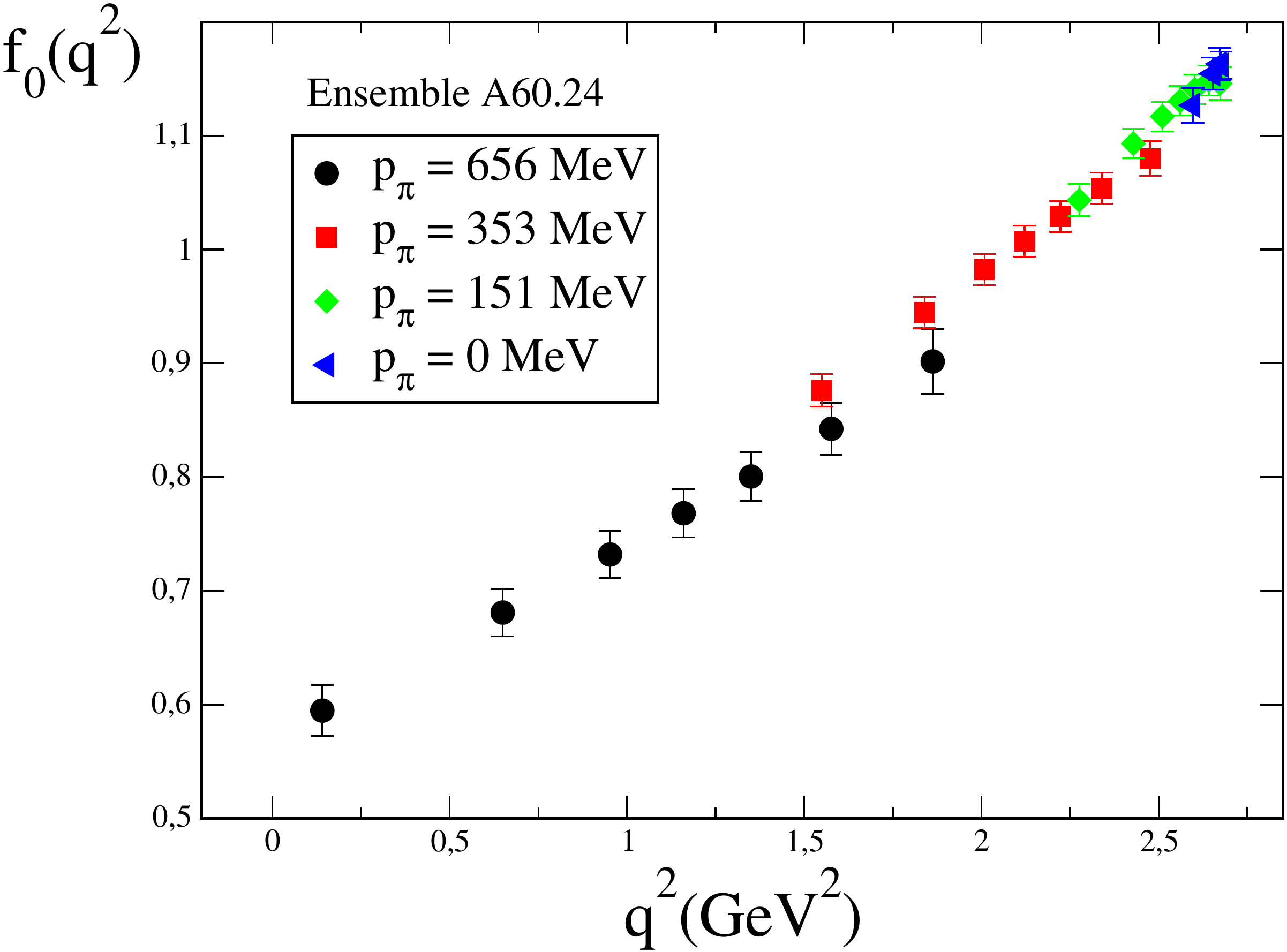}
\caption{\footnotesize \it As in Fig.~\protect\ref{fishbone}, but after the subtraction of the hypercubic effects determined by the global fit.}
\label{corrected}
\end{figure}

Using all the ingredients described above, we have performed a global fit of all the data corresponding to the time and spacial components of the vector current and to the scalar current for all the ETMC gauge ensembles, studying simultaneously the dependence on $q^2$, $m_\ell$ and $a^2$ of the Lorentz-invariant form factors $f_{+,0}(q^2)$ as well as the $q^2$ and $m_\ell$ dependence of the five hypercubic form factors $H_i$ ($i=1,...,5$). 
The quality of the fit is quite good obtaining a $\chi^2 / \rm{dof}$ equal to $\simeq 1.2$ over more than one thousands lattice points. 
Here we limit ourselves to illustrate in Fig.~\ref{corrected} the removal of the hypercubic effects, as determined by the global fitting procedure, in the case of the gauge ensemble A60.24 \cite{Carrasco:2014cwa}.

\begin{figure}[htb!]
\parbox{8.0cm}{~~ \includegraphics[scale=0.30]{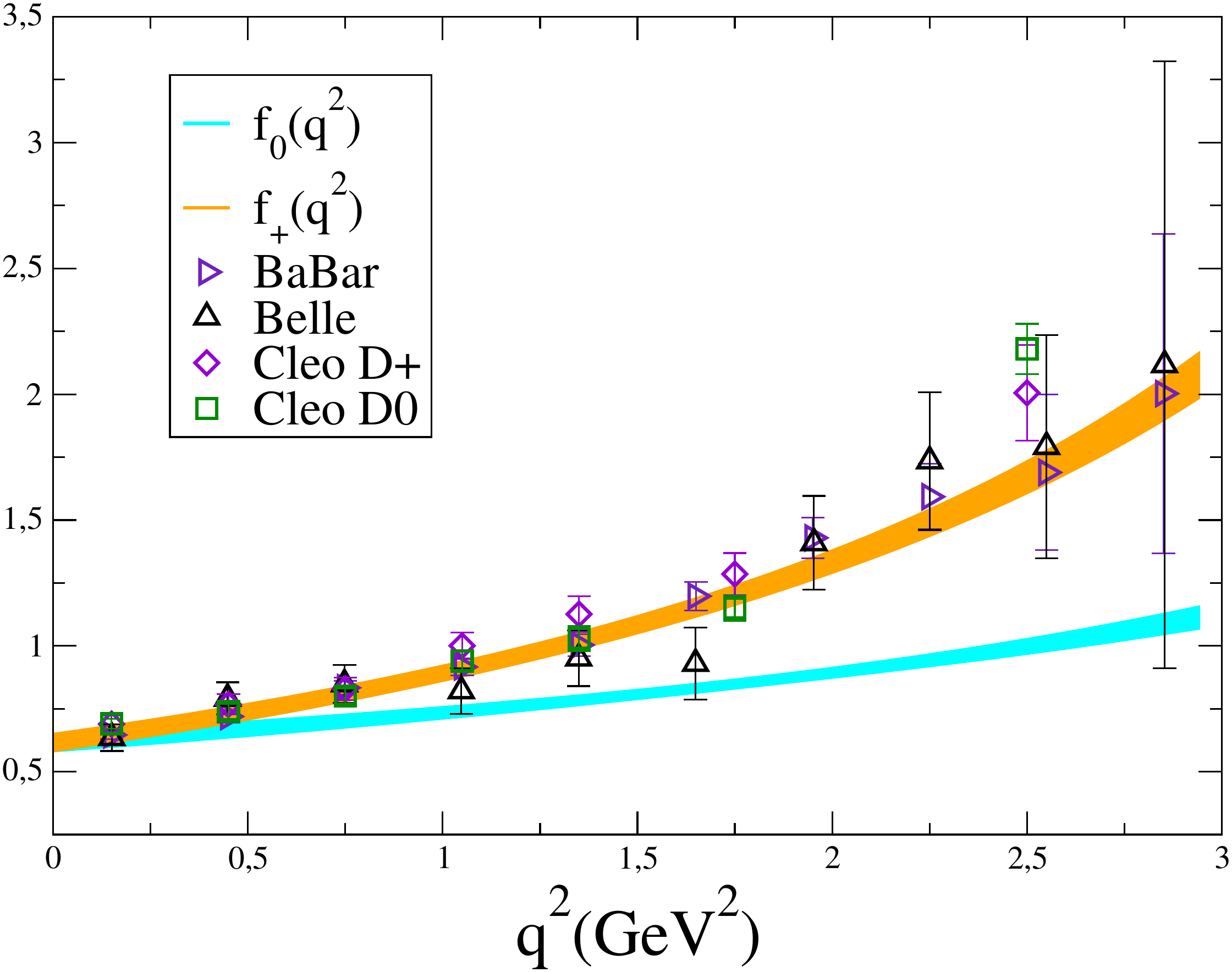}}$~$\parbox{7.0cm}{\caption{\footnotesize \it Results for the vector (orange band) and scalar (cyan band) form factors of the $D \to \pi \ell \nu$ decay extrapolated to the physical point and including the uncertainties related to the statistics, the fitting procedure, the chiral extrapolation and the continuum and infinite volume limits. For comparison the experimental data for $f_+(q^2)$ obtained by Belle \protect\cite{Widhalm:2006wz}, Babar \protect\cite{Lees:2014ihu} and Cleo \protect\cite{Dobbs:2007aa,Besson:2009uv} are shown by the different markers.}
\label{dati_sperimentali}
}
\end{figure}

In Fig.~\ref{dati_sperimentali} the momentum dependence of the vector and scalar form factors $f_+(q^2)$ and $f_0(q^2)$ extrapolated to the physical point is shown in the whole range of values of $q^2$ accessible to the experiments.
In the case of the vector form factor our results are in good agreement with the experimental data obtained by Belle \cite{Widhalm:2006wz}, Babar \cite{Lees:2014ihu} and Cleo \cite{Dobbs:2007aa,Besson:2009uv}.

Our preliminary results for the vector form factor at $q^2=0$, including the error budget, are
 \be
    f_+^{D \to \pi}(0) = 0.631 ~ (37)_{\rm stat} ~ (14)_{\rm chiral} ~ (08)_{\rm disc} = 0.631 ~ (40) ~ ,
    \label{vec_form_factor}
 \ee
which can be compared with the FLAG average $f_+^{D \to \pi}(0) = 0.666 (29)$ \cite{FLAG}, based on the lattice result obtained at $N_f = 2 + 1$ in Ref.~\cite{Na:2011mc}.

Though the presence of hypercubic effects is not new to lattice calculations, it is the first time that they have been observed in the semileptonic form factors. 
The novelty of our analysis with respect to previous studies of the semileptonic $D \to \pi $ form factors is the use of a plenty of kinematical configurations corresponding to parent and child mesons either moving or at rest. 
The use of a limited number of kinematical conditions, for instance the Breit-frame ($\vec{p}_D = - \vec{p}_\pi$) or the D-meson at rest, can mask the presence of hypercubic effects and lead to a systematic error in the extraction of the momentum dependence of the physical form factors.

\section{$\bar{D}^0-D^0$ mixing and bag-parameters}
\label{sec:DDbar}

$\Delta C=2$ transitions occurring in $\bar{D}^0-D^0$ oscillations are of special interest being the only SM process in which mixing involves up-type quarks. CP violation through these mixings is expected to be strongly suppressed within the SM, because they are dominated by light ($d$, $s$) quark exchange entailing also important long range interactions. Thus any experimental signal of CP violation in the neutral D meson sector would be a strong indication for the existence of NP~\cite{Blaylock:1995ay, Petrov:2006nc, Golowich:2007ka, Gedalia:2009kh, Ciuchini:2007cw}. Even in the absence of CP-violation, our determination of $\Delta C=2$ operator matrix elements allows to put constraints on NP models.
We compute on the lattice meson-anti-meson matrix elements of the whole basis of dimension-six four-fermion operators contributing the most general form of the effective $\Delta F=2$ Hamiltonian~\cite{Beall:1981ze, Gabbiani:1996hi,  Gabbiani:1988rb, 
Gabrielli:1995bd} 
\begin{equation}
\langle \bar{D}^0 | {\cal H}_{\textrm{eff}}^{\Delta F=2} | D^0 \rangle=\frac{G_F M^2_W}{16\pi^2}{\displaystyle \sum_{i=1}^{5}C_{i}(\mu)\langle \bar{D}^0 | \widehat {\cal O}_{i}(\mu)| D^0 \rangle} \, ,
\label{eq:Heff}
\end{equation}
where $C_i$ are the Wilson coefficients that describe short distance effects. Accordingly, they will also depend on the heavy degrees of freedom possibly circulating in loops.
In Eq.~(\ref{eq:Heff}), beyond the ``left-left'' operator relevant for the SM, flavor-changing extra terms appear: 
{\renewcommand{\arraystretch}{1.3}
\begin{equation} \label{eq:operators}
\begin{array}{ll}
{{\cal O}_{1}=\left[\bar{h}^{\alpha}\gamma_{\mu}(1-\gamma_{5}){\ell}^{\alpha}\right]\left[\bar{h}^{\beta}\gamma_{\mu}(1-\gamma_{5}){\ell}^{\beta}\right],} &
{\cal O}_{2}=\left[\bar{h}^{\alpha}(1-\gamma_{5}){\ell}^{\alpha}\right]\left[\bar{h}^{\beta}(1-\gamma_{5}){\ell}^{\beta}\right] ,  \\
{\cal O}_{3}=\left[\bar{h}^{\alpha}(1-\gamma_{5}){\ell}^{\beta}\right]\left[\bar{h}^{\beta}(1-\gamma_{5}){\ell}^{\alpha}\right]\, , & 
{\cal O}_{4}=\left[\bar{h}^{\alpha}(1-\gamma_{5}){\ell}^{\alpha}\right]\left[\bar{h}^{\beta}(1+\gamma_{5}){\ell}^{\beta}\right]\, , \\
{\cal O}_{5}=\left[\bar{h}^{\alpha}(1-\gamma_{5}){\ell}^{\beta}\right]\left[\bar{h}^{\beta}(1+\gamma_{5}){\ell}^{\alpha}\right]\, .  \\
\end{array}
\end{equation}
Their matrix elements are typically expressed in terms of the bag-parameters $B_i$: dimensionless quantities defined as the ratio of the non-perturbatively computed four-fermion matrix element over the value this matrix element takes in the vacuum insertion approximation (VIA). The reason for working with ratios is that they offer the advantage of a substantial cancellation of systematic and statistical uncertainties between the numerator and the denominator. 
Once bag-parameters have been calculated, the full matrix elements of operators~\ref{eq:operators} can be reconstructed using 
\begin{equation}
\begin{array}{l}
\langle\overline{D}^{0}|{\cal O}_{1}(\mu)|D^{0}\rangle= \xi_1 B_{1}(\mu)\, m_{D^0}^{2}f_{D^0}^{2} \,\\
\langle\overline{D}^{0}|{\cal O}_{i}(\mu)|D^{0}\rangle= \xi_{i} B_{i}(\mu)\, 
\dfrac{m_{D^0}^{4} f_{D^0}^{2}}{\left(m_{\ell}(\mu)+m_{h}(\mu)\right)^{2}} 
~~~ {\mbox{for}}~~ i =2, \ldots, 5 \, ,
\label{eq:Bi-def}
\end{array}
\end{equation}
where in the r.h.s. the matrix elements obtained in VIA are expressed in terms of the $D^0$ meson mass $m_{D^0}$ and decay constant $f_{D^0}$, and where $\xi_i=\{8/3, -5/3,1/3, 2, 2/3\}$.

Estimators for the bare bag-parameters are extracted from the asymptotic time 
behaviour of the ratios of the three- to two-point correlators which for large time separations, tend to the desired $B_i$ 
\bea
\label{eq:bareB1}
\hspace{-3.cm}{\cal R}_{1}(x_0) &=& \dfrac{C_{1}(x_{0})}{C_{AP}(x_{0})C'_{AP}(x_{0})} \, \mathop {\xrightarrow{\hspace*{1.2cm}}} \limits^{y_0 \ll x_0 \ll y_0+T_{sep}} 
\,   \dfrac{\langle \overline{D}^0 | {\cal O}_{1} | D^0 \rangle}{\langle \overline{D}^0 | A_0^{h \ell} | 0 \rangle \, 
\langle 0 | A_0^{h' \ell'} | D^0 \rangle}  \equiv  B_1\, \\
\label{eq:bareBi}
{\cal R}_{i}(x_0)&=& \dfrac{C_{i=2, \ldots, 5}(x_{0})}{C_{PP}(x_{0})C'_{PP}(x_{0})} \, \mathop {\xrightarrow{\hspace*{1.2cm}}} \limits^{y_0 \ll x_0 \ll y_0+T_{sep}} 
\,  \dfrac{\langle \overline{D}^0 | {\cal O}_{i} | D^0 \rangle}{\langle \overline{D}^0 | P^{h \ell} | 0 \rangle \, 
\langle 0 | P^{h' \ell'} | D^0 \rangle} \equiv  B_i \, ,\quad i=2, \ldots, 5
\eea
Explicitly the correlators are given by
\begin{eqnarray}
\hspace{-1.5cm}&&C_i(x_0) = \sum_{\vec x} \langle 0 |
{\cal P}^{\ell' h'}(y_0 + T_{sep}) \,{\cal O}_{i} (\vec x,x_0) \, {\cal P}^{\ell h}(y_0) | 0 \rangle\, , \quad i=1, \ldots, 
5\, ,
\label{PQPi-correl}\\
\hspace{-1.5cm}&&C_{XP}(x_0) = \sum_{\vec x} 
\langle 0 | X^{h \ell}(\vec x,x_0) \,{\cal P}^{\ell h}(y_0) | 0 \rangle\, ,
\label{PP-12}\\
\hspace{-1.5cm}&&C_{XP}^\prime(x_0) = \sum_{\vec x} \langle 0 | {\cal P}^{\ell' h'}(y_0 + T_{sep})  
\,X^{h'\ell'}(\vec x,x_0)| 0 \rangle \, ,
\label{PP-34}
\end{eqnarray}
where $X$ can be either the axial current, $A_{0}$, or the pseudoscalar density, $P$. The operator ${\cal P}^{\ell h}(y_0) $ (~\mbox{${\cal P}^{\ell' h'}(y_0 + T_{sep})$}~) which creates the meson in $y_0$ (~$y_0 + T_{sep}$~), is defined as:
\begin{equation}
\label{K-WALL}
{\cal P}^{\ell h}(y_0) =   
 \sum_{\vec y} \bar q_{\ell}(\vec y, y_0) \gamma_5 q_{h}(\vec y , y_0)\, .
\end{equation}

The RCs of both bilinear and four-fermion operators have been computed non-perturbatively in the RI$'$-MOM. The calculation is presented in~\cite{Carrasco:2014cwa} and in the Appendices of~\cite{Carrasco:2015pra}. 
At each value of the simulated light quark mass our estimates of the bag-parameters are linearly interpolated to the physical charm quark mass. A simultaneous chiral and lattice spacing extrapolation to the physical value of the pion mass and to the continuum limit is finally performed. Chiral extrapolation have been performed either assuming a simple linear dependence in the light quark mass or a NLO HMChPT fit formula \cite{Becirevic:2006me}. The results are shown in Fig~\ref{bagparameterslimits}
\begin{figure}[htb!]
\centering
\includegraphics[scale=0.65]{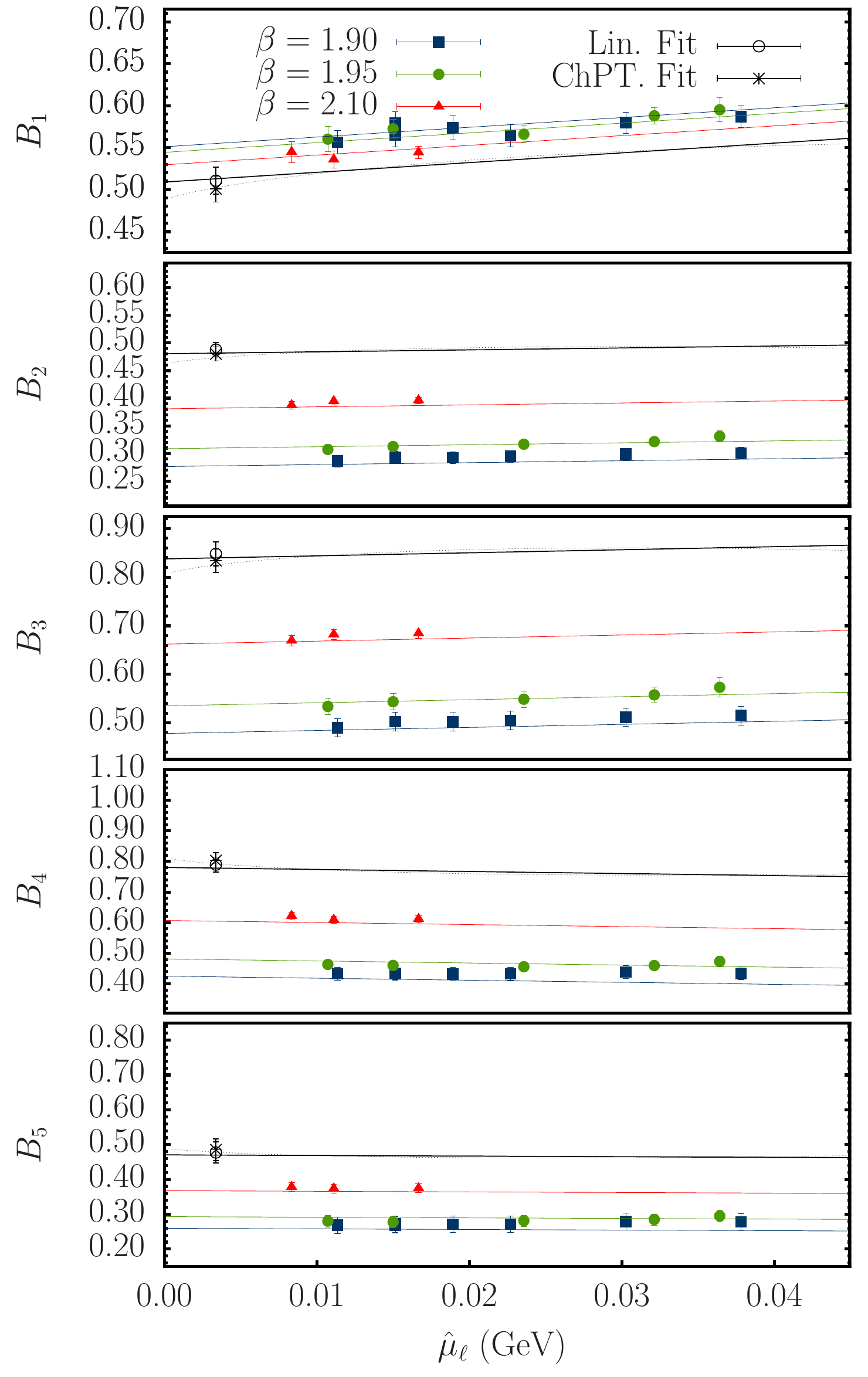}
\caption{\footnotesize \it Combined chiral and continuum extrapolation for the five $\overline{D}^0-D^{0}$ bag-parameters, $B_i$, renormalized in the $\overline{\rm{MS}}$ scheme at the scale of 3 GeV. In each panel open circles and stars represent the value at the physical point corresponding to the linear and NLO HMChPT fit, respectively.}
\label{bagparameterslimits}
\end{figure}

In Table~\ref{tab:Bi-D-all} we summarize the results for the bag-parameters 
relevant for the case of the $\overline{D}^0-D^{0}$ oscillations. 
For results given in the $\overline{\rm{MS}}$ scheme the second error we quote represents
our estimate of the systematic uncertainty coming from the perturbative matching between RI$'$ and $\overline{\rm{MS}}$ schemes. 
The uncertainties stemming only from our lattice computations are given by the first quoted error and
range from about 4\% to 8\%.
{\renewcommand{\arraystretch}{1.3}
\begin{table}[!ht]
\begin{centering}
\begin{tabular}{|c c c c c c|}
\hline 
\multicolumn{6}{|c|}{$\overline{D}^0-D^0$}\tabularnewline
\hline 
\hline
 $\overline{\rm{MS}}$ (3 GeV) & 0.757(27)(4)& 0.65(3)(2)& 0.96(8)(2) & 0.91(5)(4) & 0.97(7)(1) \tabularnewline
\hline
RI$'$ (3 GeV)                & 0.744(27) &0.87(5) & 1.34(11) & 1.14(6) & 1.39(9) \tabularnewline
\hline
\end{tabular}
\par\end{centering}
\caption{\label{tab:Bi-D-all} \footnotesize \it  Continuum limit results for the bag-parameters $B_{i}$ ($i=1, \ldots, 5$) relevant to the  $\overline{D}^0-D^{0}$ mixing renormalized in the $\overline{\rm{MS}}$ scheme and in the RI$'$ scheme at the scale of $\mu= 3$ GeV. For results given in the $\overline{\rm{MS}}$ scheme the second error indicates an estimate for the systematic uncertainty coming from the perturbative matching of RI$'$ and $\overline{\rm{MS}}$ schemes. }
\end{table} 

Using the recent averages of $D$-meson mixing data derived by the UTfit collaboration \cite{Bevan:2014tha}, the imaginary part of the $D$ mixing amplitude can be very strongly constrained, leading to very tight bounds on possible CP-violating NP contributions to the mixing. 
The results presented in Tab~\ref{tab:Bi-D-all} and the ETMC results published in \cite{Carrasco:2014uya} for $N_f=2$, are compatible among themselves with similar total uncertainties. Therefore the main conclusions presented in \cite{Carrasco:2014uya} concerning model-independent constraints on the NP scale remain unchanged.
To obtain these constraints we considered the most general effective weak Hamiltonian for D mixing of dimension six operators parameterized in terms of Wilson coefficients of the form
\begin{equation}
C_i(\Lambda)=\frac{F_iL_i}{\Lambda^2}\, , \quad i=1, \ldots, 5\, ,
\end{equation}
where $F_i$ is the (generally complex) NP flavor coupling, $L_i$ is a loop factor which depends on the interactions that generate $C_i(\Lambda)$, and 
$\Lambda$ is the NP scale, i.e. the typical mass of new particles mediating $\Delta C = 2$ transitions. For a generic strongly interacting theory with an 
unconstrained flavor structure, one expects $F_i \sim L_i\sim 1$, so that the phenomenologically allowed range for each of the Wilson coefficients can be 
immediately translated into a lower bound on $\Lambda$. Specific assumptions on the flavor structure of NP correspond to special choices of the $F_i$ 
functions.
The results for the upper bounds on the immaginary part of the Wilson coefficients and the corresponding lower bound on $\Lambda$, obtained assuming $F_i = L_i= 1$, are shown in Fig~\ref{NPbounds}. Two other interesting possibilities are given by loop-mediated NP contributions proportional to either $\alpha_s^2$ or $\alpha_W^2$ in which the bounds we quote are simply downscaled by a factor $\alpha_s(\Lambda)\sim 0.1$ or $\alpha_W(\Lambda)\sim 0.03$.

\begin{figure}[htb!]
\centering
\includegraphics[scale=1.00]{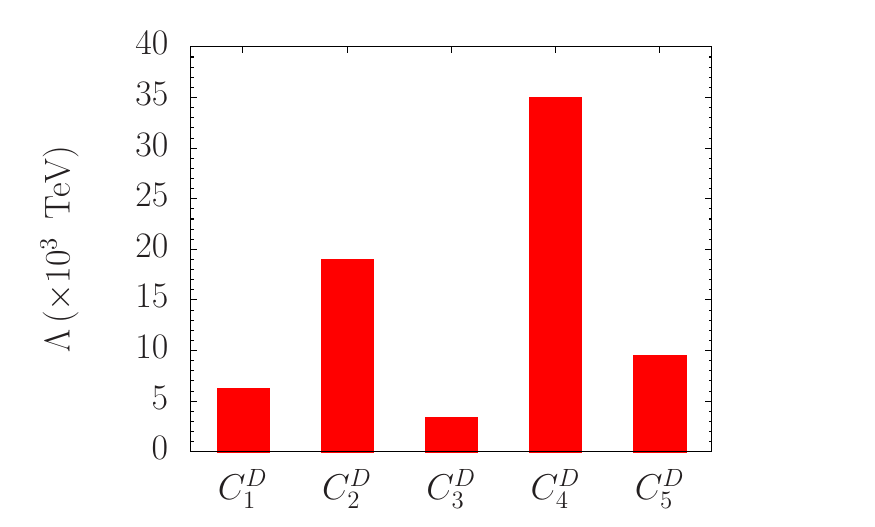}
\caption{\footnotesize \it Lower bounds on the NP scale as obtained from the constraints on the imaginary part of the Wilson coefficients.}
\label{NPbounds}
\end{figure}

In comparison with the analyses in Refs.~\cite{Bertone:2012cu, Carrasco:2013zta} dealing with neutral kaon and B-meson oscillations, we confirm that the most stringent constraints on the NP scale come from the $\overline{K}^0-K^{0}$ matrix elements, while the bounds coming from $\overline{D}^0-D^{0}$ are more stringent than those coming from $\overline{B}^0-B^{0}$.

\section{Acknowledgements}
\footnotesize{We acknowledge the CPU time provided by PRACE under the project PRA067 on the BG/Q systems Juqueen at JSC (Germany) and Fermi at CINECA (Italy), and by the agreement between INFN and CINECA under the initiative INFN-LQCD123. L. R. thanks INFN for the support under the SUMA computing project (https://web2.infn.it/SUMA).}

\end{document}